\documentclass[
 reprint,
preprintnumbers,
nofootinbib,
 amsmath,amssymb,
 aps,
]{revtex4-2}
\usepackage{graphicx}% Include figure files
\usepackage{dcolumn}% Align table columns on decimal point
\usepackage{bm}% bold math
\usepackage{xcolor} 
\usepackage{gensymb}

\begin{document}

\title{Magnetic monopoles as probes of the global structure of the Standard Model}% Force line breaks with \\

\author{Yunji Ha}
 \email{yunji.ha@durham.ac.uk}
\affiliation{%
Institute of Particle Physics Phenomenology, Department of Physics, Durham University, Durham
DH1 3LE, U.K.\\
}%

\date{\today}% It is always \today, today,

\begin{abstract}
The magnetic charges of monopoles arising in ultraviolet completions of the Standard Model are constrained by the global structure of the gauge group. After electroweak symmetry breaking, a subset of the ultraviolet monopoles carrying magnetic charges 3$g_D$ and $6g_D$ can survive as isolated, colour-neutral states in the infrared. We show that this selection rule follows from the interplay between the symmetry structure and the magnetic 1-form symmetry, and discuss how the relic abundance of such monopoles can be naturally suppressed by the symmetry breaking scale. We further demonstrate that such monopoles with Lorentz factor $\gamma\lesssim 10^4$ propagate through matter with ionisation energy loss profiles scaling as $g^2$ using CORSIKA Monte Carlo simulations. 
%\begin{description}
%\end{description}
\end{abstract}

\maketitle

%\tableofcontents
\preprint{IPPP/25/90}
%%%%%%%%%%%%%%%%%%%%%%%%%%%%%%%%%%%%%%%%%%%%%%%%%%%%%%%%%%%%%%%%%%%%%%%%%%%%%%
\section{Introduction}
In the Standard Model (SM), perturbative interactions of gauge bosons are determined by the Lie algebra of the gauge group, while non-perturbative objects such as magnetic monopoles are sensitive to its global structure. This structure is encoded in a discrete quotient group of the SM gauge group,
\begin{align}
    G_p = \frac{G_{\mathrm{SM}}}{Z_p} = \frac{\mathrm{SU(3)}_c \times \mathrm{SU(2)}_L \times \mathrm{U(1)}_Y}{Z_p}\,\,,
    \label{eq: GSM}
\end{align}
where $p$ can take values of $p=1,2,3,6$. The choice of quotient specifies how the centres of the non-Abelian factors and the phase of the hypercharge group are identified. Consistency of any ultraviolet (UV) completion then requires all matter fields to transform trivially under the corresponding centre symmetry~\cite{Hucks:1990nw,Tong:2017oea,Alonso:2024pmq,Alonso:2025rkk}. Equivalently, for a fermion field $\psi$ and a generating element of the discrete group $\xi \in Z_p$ follows $\psi \xi = \xi \psi = \psi$, which imposes non-trivial constraints on the allowed spectra of electric and magnetic charges. The fundamental homotopy of the SM gauge group by itself is insensitive to the choice of the discrete quotient $Z_p$. This degeneracy is resolved by introducing electric and magnetic 1-form symmetries, combined with the centre symmetry of the quotient choice, which provides a systematic classification of monopole fluxes and identifies the UV embedding to which they belong. After electroweak symmetry breaking (EWSB), the global structure of the unbroken theory changes and monopoles originating from UV completions realign their magnetic fluxes with the electroweak gauge basis. These new configurations are not created during the EWSB, but they are infrared (IR) descendants of pre-existing UV monopoles whose magnetic fluxes are modified into combinations compatible with the surviving gauge symmetry~\cite{Khoze:2024hlb}. It has been shown that this realignment selects the ones among allowed UV monopoles that are free of chromomagnetic charge and therefore remain colourless in IR. There are several complementary ways to experimentally probe this global structure of the SM. Since different quotient choices lead to distinct spectra of allowed electric charges, one possibility is the search for fractionally charged particles, either at colliders or through cosmological and astrophysical signatures~\cite{Koren:2024xof, Langacker:2011db}. Consequently, searches for magnetic monopoles provide a probe for the SM global structure through the Dirac quantisation condition. The existence of dynamical monopoles or fractionally charged states would explicitly break electric or magnetic 1-form symmetries in the UV, while their non-observation at accessible energy scales points towards an emergent 1-form symmetry in the IR. Observation of magnetic monopoles will thereby constrain the viable UV embeddings of the SM. We perform a quantitative comparison between the analytic expectation and full Monte Carlo (MC) simulation of atmospheric monopole cascades using the CORSIKA framework combined with CONEX~\cite{Heck:1998vt}, confirming the expected charge scaling. The rest of this paper is organised as follows. In Section II, we analyse how the global structure of the SM gauge group constrains the magnetic flux sectors of UV monopoles after EWSB. Section III discusses the implications of quotient choices for monopole production and relic abundances. In Section IV, we study monopole interactions with matter, review the relevant energy loss formula, and present the MC simulation results. Section V summarises our conclusions and outlines phenomenological consequences and prospects for detectors. 

%%%%%%%%%%%%%%%%%%%%%%%%%%%%%%%%%%%%%%%%%%%%%%%%%%%%%%%%%%%%%%%%%%%%%%%%%%%%%%

\section{Magnetic fluxes before and after EWSB}

To determine which magnetic flux sectors can arise from UV monopoles and survive after EWSB, we first recall how the global structure of the SM in eq.~(\ref{eq: GSM}) constrains the allowed electric and magnetic charges. These constraints are encoded in the Dirac quantisation condition for electric charge $q_e$ and magnetic charge $q_m$, embedded in the Cartan subalgebra, parameterised by a set of integers $\{ n_1,n_2,m,q_m\}$, corresponding to the diagonal generators of the non-Abelian factors and the hypercharge of $\mathrm{U(1)}_Y$. In this parameterisation, the Dirac quantisation condition takes the form 
\begin{align}
    &\exp\Bigg[{2\pi i \sum^{r}_{i=1}q^i_eq^i_m T^i}\Bigg] =1
    \label{eq:smdirac}
\end{align}
which may be explicitly written in terms of the SM Cartan generators 
\begin{align}
     &\exp\bigg[{2\pi i \bigg(-n_c \frac{N}{3} + n_L \frac{m}{2} + Q_Yq_m\bigg)}\bigg] = 1\,\,.
     \label{eq:newDirac}
\end{align}
Here $N = n_1 + n_2$, $n_c$ is the triality defined mod 3 (with $n_c = 1$ corresponding to the fundamental representation of $\mathrm{SU(3)}_c$), and $n_L$ is the duality index of $\mathrm{SU(2)}_L$, defined mod 2, for which the fundamental representation has $n_L = 1$. Magnetic fluxes quantised according to eq.~(\ref{eq:newDirac}) specify all 't~Hooft line operators consistent with the SM charge assignments and the representations of chosen quotient group. At distances larger than the non-Abelian monopole core, any dynamical monopole realising such a flux sector reduces to an effective Dirac monopole configuration. We label the six distinct magnetic flux sectors by $M_i$, with $i=1,\dots,6$. Different UV embeddings allow different subsets of these flux sectors, and fix the minimal magnetic charge carried by dynamical monopoles. For example, the $G_{\mathrm{SM}}/Z_6$ quotient, as realised in an $\mathrm{SU(5)}$ embedding, allows all six $M_i$ monopoles with the minimal magnetic charge equal to one unit of the Dirac charge $g_D$. However, a $G_{\mathrm{SM}}/Z_3$ embedding, such as the $\mathrm{SU(4)}_c \times \mathrm{SU(2)}_L \times \mathrm{SU(2)}_R$ Pati-Salam group only permits the $M_2$, $M_4$ and $M_6$ sectors, corresponding to magnetic charges $2g_D$, $4g_D$ and $6g_D$, respectively. The $G_{\mathrm{SM}}/Z_2$ quotient admits the $M_3$ and $M_6$ flux sectors, corresponding to magnetic charges $3g_D$ and $6g_D$, and can be realised in a $\mathrm{SU(3)}_c \times \mathrm{SU(3)}_L \times \mathrm{SU(3)}_R$ embedding. Finally the trivial quotient $G_{\mathrm{SM}}/Z_1$, which may be embedded in the non-Abelian gauge group $\mathrm{SU(3)}_c \times \mathrm{SU(2)}_L \times \mathrm{SU(2)}_Y$ (with $\mathrm{U(1)}_Y$ embedded in $\mathrm{SU(2)_Y}$), only admits the $M_6$ monopoles with minimal magnetic charge $q_m = 6 g_D$~\cite{Alonso:2025rkk}.

EWSB determines the unbroken gauge group and modifies the global structure inherited from its UV realisation. Before EWSB, the relevant centre symmetry acting on line operators contains a $Z_6$ subgroup, which is isomorphic to $Z_3 \times Z_2$. The $Z_2$ factor is associated with the centre of $\mathrm{SU(2)}_L$ of the broken symmetry, and spontaneous symmetry breaking leaves $Z_3$ (or trivial) centre in the unbroken gauge group. Explicitly, the unbroken gauge group after EWSB takes the form
\begin{align}
   G_p \;=\; \frac{G_{\mathrm{SM}}}{Z_p}
   \;\longrightarrow\;
   \frac{\mathrm{SU(3)}_c \times \mathrm{U(1)}_{\mathrm{em}}}{Z_{p'}} ,
   \qquad p' = 1,3 .
   \label{eq:ewsb}
\end{align}
where $p'$ labels the quotient of the unbroken gauge group after the symmetry breaking. As a consequence of EWSB, the magnetic flux sectors of UV monopoles realign and are projected onto the electroweak gauge basis. In particular, the magnetic fluxes are characterised by $ \{\pm n_1, \pm n_2, \pm (n_1 - n_2)\}$ that takes values of $0$ or $\pm 1$. See \cite{Alonso:2025rkk, Khoze:2024hlb} for detailed analysis and derivation.

Among the UV monopoles, the $M_3$ and $M_6$ sectors are realised for $p=2$ and $p=6$, respectively. After EWSB, these correspond to $p'=1$ and $p'=3$, giving rise to the IR descendants $M'_3$ and $M'_6$ at the electroweak scale. These monopoles are distinguished by the absence of chromomagnetic charge, which implies that they are not attached to confining colour strings and therefore survive as genuine IR degrees of freedom. The $M'_6$ monopole can be embedded in all UV gauge groups $G_1$, $G_2$, $G_3$ and $G_6$ and is the minimum allowed magnetic charge when it is from $G_1$. However, unless the embedding theory is $G_6$, a monopole with magnetic charge $q_m=6g_D$ is not fundamental but arises as a composite configuration. The $M'_3$ monopole can be embedded in $G_2$ and $G_6$, and is fundamental only when the UV theory is $G_2$. 

We can reformulate this statement using a higher-form symmetry argument. Although the unbroken gauge group after electroweak symmetry breaking admits non-trivial magnetic line operators associated with 't Hooft lines, the fundamental homotopy group alone is insensitive to the discrete quotient data $p$ and $p'$.  A classification of the allowed magnetic flux sectors sensitive to the UV completion of the theory requires the introduction of electric and magnetic 1-form symmetries. The unbroken electromagnetic field strength after electroweak symmetry breaking is given by 
\begin{align}
\begin{aligned}
      &F^{\rm em}_{\mu\nu}
= \sin\theta_W\,B_{\mu\nu}
+ \cos\theta_W\,W^3_{\mu\nu},\\
&\widetilde F^{\rm em}_{\mu\nu}
= \tfrac12\epsilon_{\mu\nu\rho\sigma}F^{\rho\sigma}_{\rm em}.
\end{aligned}
\end{align}
where $B_{\mu\nu}$ and $W^3_{\mu\nu}$ are the field strengths of $\mathrm{U(1)}_Y$ and the neutral $\mathrm{SU(2)}_L$ gauge field respectively. In the monopole background, far outside the non-Abelian core, the long-distance electromagnetic fields take the form
\begin{align}
    B^{\rm em}_i \equiv \tilde F^{0i}_{\rm em}
= g_{\rm em}\,\frac{x^i}{4\pi r^3}.
\end{align}
To characterise the magnetic charge compatible with the quotient structure of the unbroken gauge group, we introduce the operator $Q_3$. Following the convention of~\cite{Alonso:2025rkk}, $Q_3$ is a symmetry generator defined by
\begin{align}
  Q_3 \equiv 3\,(1 + p' k')\, Q_{\rm em} + \tilde{\lambda}_8 ,
\end{align}
where $Q_{\rm em}$ denotes the electromagnetic charge operator and $k'$ parametrises the scaling of the hypercharge, often referred to as the \emph{compositeness degree}~\cite{Alonso:2024pmq}.
The spectrum of $Q_3$ forms a discrete lattice characterised by
\begin{equation}
  Q_3 \in p'\mathbb{Z}.
\end{equation}
Denoting the eigenvalues of $Q_3$ by $q_3$, one may introduce the index $n_3$ defined modulo~3, such that $n_3 = q_3 \bmod 3$.
The action of the discrete symmetry on a state $|q_3\rangle$ is then given by
\begin{align}
  \begin{aligned}
      &e^{2\pi i \ell Q_3/p'} |q_3\rangle
  = e^{2\pi i \ell n_3/p'} |q_3\rangle\,,\\
  &n_3 \equiv q_3 \!\!\!\mod 3 .
  \end{aligned}
  \label{eq:q3def}
\end{align}
The corresponding magnetic index $n_3^m$ is defined in a manner consistent with the Dirac quantisation condition. Now the magnetic charge $g_{\mathrm{em}}$ can be written as
\begin{align}
    g_{\mathrm{em}} = \frac{n^m_3 Q_3}{3}.
\end{align}
Acting on a state with $Q_3$ eigenvalue $q_3$, the magnetic charge reduces to $g_{\mathrm{em}} = n_3^m q_3/3$. Using Gauss’s theorem, the magnetic field satisfies
\begin{equation}
  \partial_i B^i_{\rm em}
  = g_{\rm em}\delta^{(3)}(x),
\end{equation}
which identifies the monopole as a point-like magnetic source. It then follows that
\begin{equation}
  \partial_\mu \widetilde F^{\mu 0}_{\mathrm{em}}
  = \partial_i B^i_{\rm em}
  = g_{\rm em}\delta^{(3)}(x).
\end{equation}
To describe a moving monopole and its associated worldline, it is convenient to introduce the magnetic current.
Boosting the rest-frame current $\tilde{J}^\mu_{\rm em,rest}=\partial_\mu \tilde F^{\mu 0}_{\rm em}$ along the $z$-axis with velocity $\beta$ gives
\begin{equation}
  \tilde{J}^\mu_{\rm em}(x)= g_{\mathrm{em}}\gamma(1,0,0,\beta)\,
    \delta(x)\delta(y)\delta\bigl(\gamma(z-\beta t)\bigr).
\end{equation}
which corresponds to a straight monopole worldline along the $z$-direction. The magnetic 1-form symmetry $Z_{p'}^{\mathrm{mg} (1)}$ is generated by the ’t~Hooft line operator $T_{n_3^m}(C)$ associated with the monopole worldline $C$, which transforms as
\begin{align}
    Z^{\mathrm{mg}(1)}_{p'} : T_{n^m_3 (C)} \rightarrow U_{\ell m} (\Sigma_+) T_{n^m_3} (C) U^\dagger_{\ell m} (\Sigma_-).
    \label{eq:magsym}
\end{align}
To make the action of the 1-form symmetry explicit, we consider a three-dimensional volume $V_3$ whose boundary links the monopole worldline.  The enclosed surface $C$ of magnetic flux emitted by the 't Hooft line is parameterised by the volume $V_3(t,x,y,z)$, two normal vectors $(n^\perp)^\mu, (n'^\perp)^\mu$ at the boundaries and for surfaces $\Sigma_\pm$ which are the upper and the lower boundaries of the slab enclosing $C$. For a genuine magnetic monopole, the magnetic 1-form symmetry operator
$U_{\ell_m}$ acts on $V_3$ as
\begin{align}
    U_{\ell_m }((\partial V_3)_\pm) = \exp \Big( i2\pi \frac{\ell_m}{p'} \int_{\Sigma_\pm} d\Sigma_\pm n^\perp_\mu n'_\nu \tilde{F}_{\mathrm{em}}^{\mu\nu} \Big)\,,
\end{align}
following the general construction of magnetic 1-form symmetry operators~\cite{Alonso:2025rkk}. This operator measures electromagnetic flux through the surface $\Sigma_\pm$, and its exponential phase is the topological linking between a magnetic line and the surface. Performing the unitary transformation of the magnetic 1-form symmetry  
\begin{align}
    &U_{\ell m} (\Sigma_+) T_{n^m_3} (C) U^\dagger_{\ell m} (\Sigma_-)\\
    &= \exp \Big( 2\pi i \ell_m \frac{n^m_3 Q_3}{3p'} \mathrm{Link}(V_3, C) \Big)\ T_{n^m_3} (C),
\end{align}
where we used the Bianchi identity modified for presence of magnetic source $\partial_\mu \widetilde{F}^{\mu \nu }_{\mathrm{em}} = \tilde{J}^\nu $. The linking number $\mathrm{Link}(V_3,C)$ counts the number of times the monopole worldline $C$ pierces the three-dimensional volume $V_3$ and takes integer values.
In the this configuration $\mathrm{Link}(V_3,C)=1$, so after EWSB we have
\begin{align}
   T_{n^m_3}(C) \rightarrow \exp \Big[ 2\pi i \ell_m \frac{n^m_3 Q_3}{3p'} \Big] T_{n^m_3}(C) ,
\end{align}
showing that the action of the magnetic 1-form symmetry reduces to a pure phase. The transformation is trivial if and only if $n_3^m$ is a multiple of $3$, in which case the associated ’t~Hooft line is neutral under the magnetic 1-form symmetry and can be screened by dynamical excitations.
For $n_3^m\notin 3\mathbb{Z}$, the phase is non-trivial, indicating a non-vanishing magnetic 1-form charge. Explicitly, we find that EWSB enforces a selection rule,
allowing only $M'_3$ and $M'_6$ monopoles with magnetic charges $3g_D$ and $6g_D$ to survive as isolated colour-neutral states. All other monopoles have a non-trivial phase picked up by the transformation in eq.(\ref{eq:magsym}) when the magnetic flux is the fundamental (i.e., minimal non-zero) magnetic charge allowed by the given UV embedding. This shows that the surviving IR monopole species are determined by the compatibility of their magnetic charges with the electromagnetic charge lattice inherited from the UV completion. Having identified the IR monopole species, we now turn to their production mechanisms and number densities, which determine their phenomenological relevance and observational prospects.

%%%%%%%%%%%%%%%%%%%%%%%%%%%%%%%%%%%%%%%%%%%%%%%%%%%%%%%%%%%%%%%%%%%%%%%%%%%%%%
\section{Production and relic abundance of monopoles}
While the allowed magnetic charges are fixed by the global structure of the SM gauge group, the production rate and abundance of monopoles depend on the symmetry breaking scale at which the monopoles are produced. In this section, we analyse how different UV embeddings lead to distinct monopole abundances.

Magnetic monopoles can be produced during cosmological second-order phase transitions in the early universe if the vacuum manifold of breaking symmetry has a non-trivial second homotopy group~\cite{Kibble:1976sj,Kibble:1980mv,Zurek:1985qw,Preskill:1979zi}. At the time of production, causality limits the correlation length
$\xi(T_c)$ at the critical temperature $T_c$, leading to an initial monopole number density $n_M(T_c) \sim \xi^{-3}(T_c)$. After the production, initial monopoles and antimonopoles quickly annihilate and the final relic abundance reduces. This production mechanism underlies the classical $\mathrm{SU}(5)$ monopole problem. GUT monopoles produced at
$T_c \sim 10^{15}-10^{16}\,\mathrm{GeV}$ with monopoles mass
$M \sim 10^{16}\,\mathrm{GeV}$ will overclose the universe unless diluted by inflation or avoided through a strongly first-order phase transition~\cite{Preskill:1979zi}. However, UV completion models that break at much lower scales naturally produce monopoles via the Kibble mechanism with relic abundances far below cosmological bounds. 

For benchmark relic abundances, we consider the upper bound on the correlation length, $\xi(T_c)\lesssim H^{-1}(T_c)$, which yields a horizon-limited abundance
\begin{align}
    \begin{aligned}
        r_H(T_c) \equiv \frac{n_M(T_c)}{T_c^3}
\sim  \bigg(\frac{H(T_c)}{T_c} \bigg)^3
    \end{aligned}
\end{align}
in radiation dominated universe, 
\begin{align}
    \begin{aligned}
        r_H(T_c) \sim  \left(\frac{1.66\sqrt{g_*}\,\,T_c}{M_{\mathrm{Pl}}}\right)^3 .
    \end{aligned}
\end{align}
For $M_{\mathrm{Pl}}=1.22 \times 10^{19} \mathrm{GeV}$ and taking $g_* \simeq 10^2$ for simplicity, this gives
\begin{align}
r_H(T_c=10^{10}\,{\mathrm{GeV}}) &\sim 10^{-24}, \\
r_H(T_c=10^8\,{\mathrm{GeV}}) &\sim 10^{-30}, \\
r_H(T_c= 10^2\,{\mathrm{GeV}}) &\sim 10^{-48},
\end{align}
showing significant suppression associated with low-scale symmetry breaking and including monopole-antimonopole annihilation would further suppress the abundance\footnote{This estimate should be understood as order-of-magnitude benchmarks for the monopole abundance at formation.}. For example, consider $\mathrm{SU(3)}_c \times \mathrm{SU(3)}_L \times \mathrm{SU(3)}_R$, which admits $3g_D$ as a minimal non-trivial magnetic charge. Trinification models do not predict a unique symmetry breaking scale, and explicit constructions in the literature realise trinification breaking at scales ranging from GUT-scale scenarios down to intermediate or low scales~\cite{Raut:2022ryj}. For intermediate or low-scale symmetry breaking scenarios, the resulting relic abundance of $M'_3$ monopoles is many orders of magnitude below that of GUT-scale monopoles. A low-scale benchmark can also arise in embeddings with quotient parameter $p=1$, for which the final stage of symmetry breaking to the SM gauge group occurs at $T_c=\mathcal{O}(10^2)\,\mathrm{GeV}$ if the UV theory is $\mathrm{SU(3)}_c \times \mathrm{SU(2)}_L \times \mathrm{SU(2)}_Y$~\cite{Alonso:2025rkk}. In this case, the minimal magnetic charge is $g=6g_D$, corresponding to the IR monopole $M'_6$. The relic abundance of such monopoles is naturally suppressed without invoking inflation. Thus, the relic abundance of monopoles is a UV-sensitive quantity relevant to both the symmetry breaking scale and the magnetic charge lattice of the SM quotient structure. In the following section, we discuss how these non-minimal Dirac charge monopoles can possibly be tested through cosmic-ray searches.
%%%%%%%%%%%%%%%%%%%%%%%%%%%%%%%%%%%%%%%%%%%%%%%%%%%%%%%%%%%%%%%%%%%%%%%%%%%%%%

\section{Monopole-matter interaction and atmospheric propagation}
After EWSB, charged chiral fermions combine into massive Dirac fermions, and the scattering of monopoles with matter becomes better defined. Solving the Dirac equation in a monopole background yields the helicity-flip and non-flip Kazama-Yang-Goldhaber (KYG) amplitudes~\cite{Kazama:1976fm}, incorporated in the modified Bethe-Bloch energy loss formula for magnetically charged particles~\cite{Ahlen:1978jy}. 
\begin{align}
\begin{aligned}
      -\frac{dE}{dx} = \frac{4\pi N g^2 e^2}{mc^2} \bigg[ \ln \frac{2mc^2\beta^2\gamma^2}{I} + \frac{K(|g|)}{2} \\   -\frac{1}{2} -\frac{\delta}{2} -B(|g|) \bigg]\,,
\end{aligned}
    \label{eq:ahlen}
\end{align}
where \(g\) denotes the monopole magnetic charge, $N$ is the electron number density of the medium, $m$ is the electron mass, $I$ is the (magnetic) mean ionisation potential, which coincides with the usual mean excitation energy for gaseous media, $\beta$ and $\gamma$ are the monopole velocity and Lorentz factor, and $\delta$ accounts for the density effect~\cite{STERNHEIMER1984261}. $K(|g|)$ and $B(|g|)$ are the KYG and Bloch corrections, respectively, that have been appropriately scaled for monopoles with higher magnetic charges~\cite{DERKAOUI1998173}.
For ultra-relativistic monopoles traversing the atmosphere, pair-production and photo-nuclear interactions dominate the energy loss, whereas for moderately relativistic monopoles (with Lorentz factor $\gamma \lesssim 10^4$) ionisation energy loss becomes dominant, while the bremsstrahlung process remains suppressed due to the large mass of monopoles (see fig.1 in~\cite{PierreAuger:2016imq}). The resulting stopping power shows a quadratic dependence on the magnetic charge, which can be used for a quantitative detection strategy in experiments.

\begin{figure}[t]
  \centering
  \includegraphics[width=0.48\textwidth]{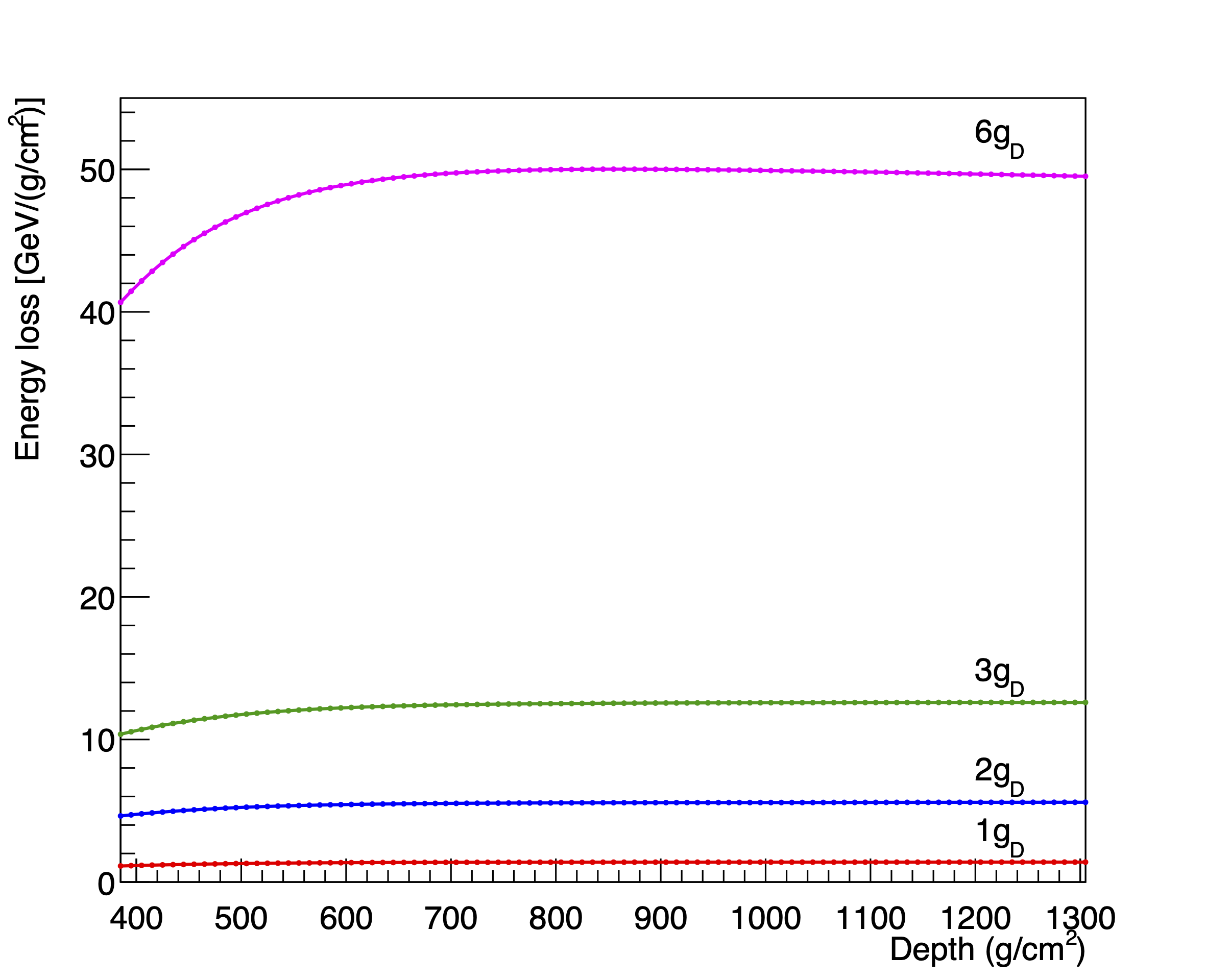}
  \includegraphics[width=0.48\textwidth]{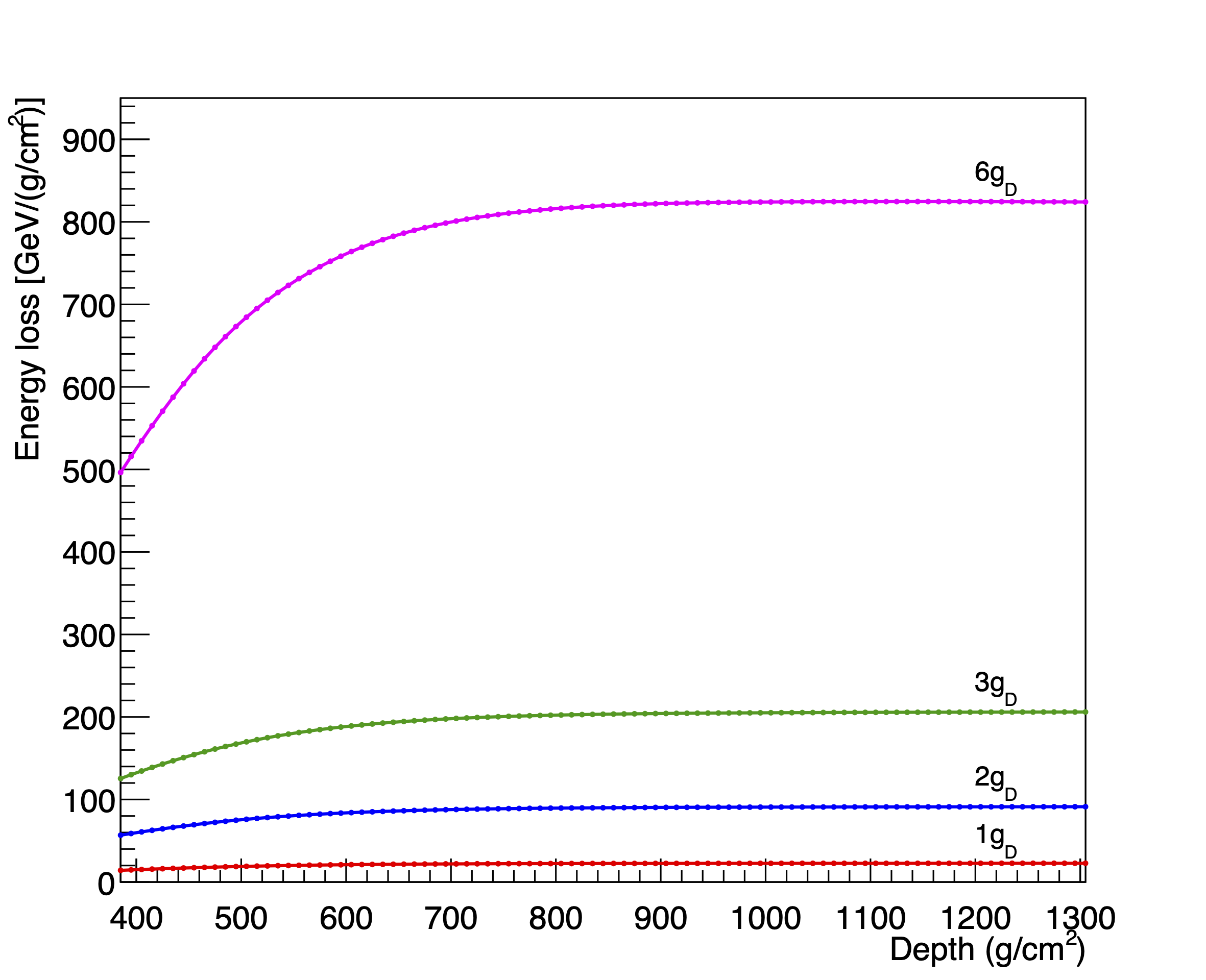}
  \caption{Average energy loss by ionisation per unit atmospheric depth for magnetic monopoles with charges $g = n g_D$ propagating through the atmosphere, shown as a function of column depth. The upper and lower panels correspond to Lorentz factors $\gamma = 10^{3}$ and $\gamma = 10^{4}$, respectively. The curves correspond to $n = 1, 2, 3, 6$. The depth range shown focuses on the region where the monopole has fully entered
the atmosphere, and the stopping power varies slowly with depth.}
  \label{fig:energyloss}
\end{figure}

To study the propagation of magnetic monopoles with intermediate mass ($10^5~\mathrm{GeV}$) through the atmosphere, we performed MC simulations using the CORSIKA framework (version 7.7550), supplemented by the CONEX cascade solver~\cite{Heck:1998vt} for efficient computation of the longitudinal energy deposition profile for magnetic monopoles. High-energy hadronic interactions were modelled using the EPOS-LHC model, while low-energy hadronic interactions were treated with UrQMD~1.3cr, as implemented in CORSIKA. The detector geometry was chosen to be a horizontal, flat detector array with a zenith angle of $70\degree$ for monopole searches at the Pierre Auger Observatory. The simulations were performed for moderately relativistic monopoles with Lorentz factors in the range $\gamma = E_{\mathrm{Mon}}/M_{\mathrm{Mon}} \sim 10^3–10^4$, for which energy loss in the atmosphere is dominated by ionisation. While the analytic solution remains applicable down to lower Lorentz factors, we restrict our numerical study to this range, where the energy-loss profile is smooth and directly relevant for air shower observables.

\begin{figure}[t]
  \centering
  \includegraphics[width=0.48\textwidth]{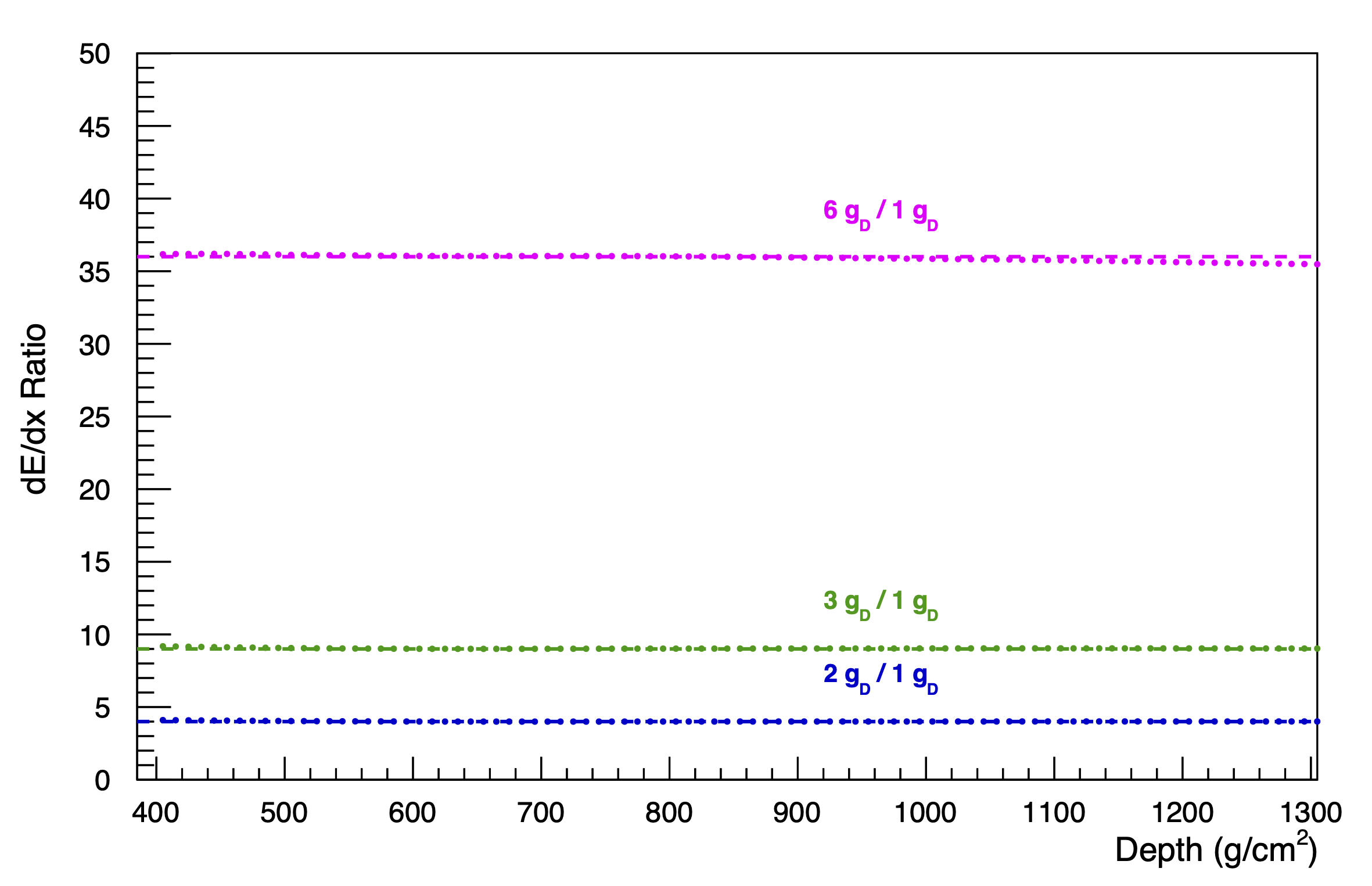}
  \includegraphics[width=0.48\textwidth]{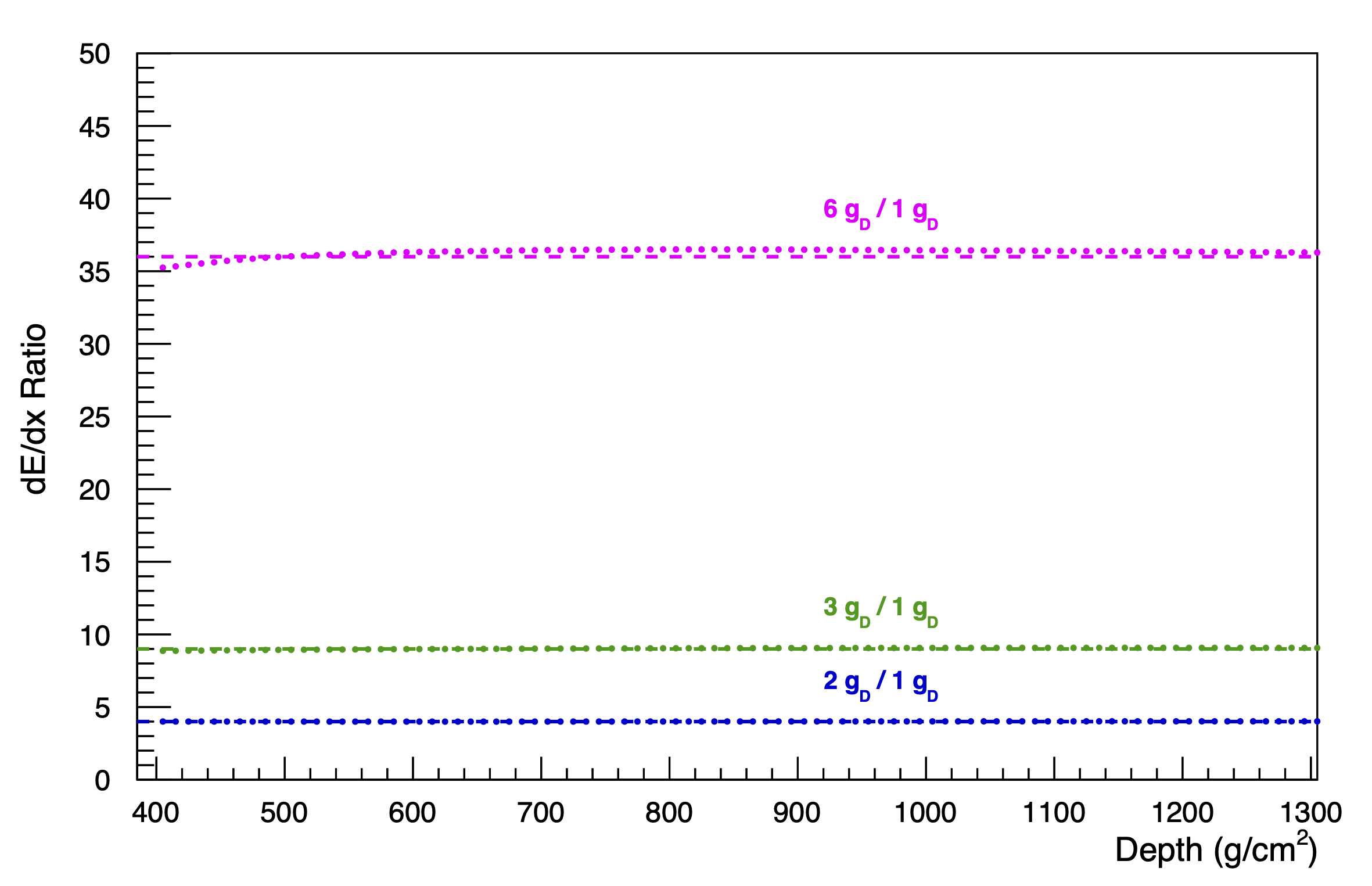}
  \caption{Ratio of the average energy loss by ionisation per unit atmospheric depth for magnetic monopoles with charge $g = n g_D$ relative to the unit Dirac charge case, shown as a function of depth. The upper and lower panels correspond to $\gamma = 10^{3}$ and $\gamma = 10^{4}$, respectively. The ratios are approximately constant and follow the expected $n^2$ (equivalently $g^2$) scaling within the ionisation-dominated velocity regime.}
  \label{fig:ratio}
\end{figure}

The MC simulations demonstrate that monopoles, which are free of chromomagnetic flux, propagate through the atmosphere as stable, highly ionising particles with smooth longitudinal energy deposition in the simulated profiles. As shown in Secs.~II–III, the requirement of vanishing chromomagnetic flux after EWSB implies that the monopoles with magnetic charges $g=3g_D$ and $6g_D$ remain as isolated IR objects. The results of this section show that these monopoles admit a clean and experimentally distinctive atmospheric signature. Since the existence of such chromomagnetic flux free monopoles is dictated by the global structure of the SM gauge group, the observation of highly ionising atmospheric tracks of this type may provide an indirect probe of the SM quotient structure.

The values $n=1,2,3,6$ considered in the simulations correspond to the minimal magnetic charges realised in possible minimal UV embeddings of the SM gauge group. Specifically, $n=1$ arises as the minimal Dirac monopole in $\mathrm{SU(5)}$ GUT, while $n=2$ corresponds to the minimal monopole allowed in Pati-Salam embeddings. The case $n=3$ is realised as the minimal non-trivial magnetic charge in $p=2$, whereas $n=6$ corresponds to the minimal magnetic charge in embeddings with trivial quotient structure ($p=1$), that could be embedded in $\mathrm{SU(3)}_c \times \mathrm{SU(2)}_L \times \mathrm{SU(2)}_Y$ suggested in~\cite{Alonso:2025rkk}. However, magnetic charges $n=4$ and $n=5$ do not arise as fundamental objects in any consistent UV embeddings we considered. They necessarily correspond to composite configurations formed from lower-charge monopoles and are therefore excluded from the current analysis. The cases $n=1,2$ are included here for comparison and validation of the expected charge scaling.
The result shows that the average energy loss scales quadratically with the magnetic charge, $\langle dE/dX \rangle \propto g^2$, indicating a good agreement with the eq.(\ref{eq:ahlen}) as shown in fig.~\ref{fig:energyloss} and fig.~\ref{fig:ratio}

%%%%%%%%%%%%%%%%%%%%%%%%%%%%%%%%%%%%%%%%%%%%%%%%%%%%%%%%%%%%%%%%%%%%%%%%%%%%%%

\section{Conclusions}

The framework developed here provides a testable link between the global structure of the SM and magnetic monopole phenomenology. After EWSB, only a restricted subset of monopoles survives as isolated, colour-neutral objects in the IR regime. In particular, the monopoles carrying magnetic charges $3g_D$ and $6g_D$, corresponding to the IR species $M'_3$ and $M'_6$, are the only monopoles that can remain as genuine isolated degrees of freedom. Monopoles with other charges either do not arise as fundamental objects in any consistent UV embedding of the SM, or are confined by chromomagnetic fluxes. We verified that this selection rule follows directly from the global structure of the gauge group. 

We have further analysed the cosmological production of monopoles in selected UV completions. For symmetry breaking scales at intermediate or low energies, the relic abundance of monopoles produced via the Kibble mechanism is naturally suppressed, without the need to invoke inflation. As a result, the monopoles $M'_3$ and $M'_6$ emerge as rare but potentially observable relics of the UV completion of the SM.

Motivated by this observation, we studied the propagation of magnetic monopoles through matter and the atmosphere. Using MC simulations, we confirmed that for monopoles with $10^3 \leq \gamma \leq 10^4$, the energy loss by ionisation scales quadratically with the magnetic charge,
$\langle dE/dX \rangle \propto g^2$, in agreement with analytic expectations from the modified Bethe-Bloch formula.  Our results highlight $M'_3$ and $M'_6$ monopoles as distinctive probes of the global structure of the SM. The observation of highly ionising monopole signals would provide a window into the quotient structure of the SM gauge group and its UV embeddings.

%%%%%%%%%%%%%%%%%%%%%%%%%%%%%%%%%%%%%%%%%%%%%%%%%%%%%%%%%%%%%%%%%%%%%%%%%%%%%%

\acknowledgments
We would like to thank Valentin Khoze and Rodrigo Alonso for their helpful discussions. Y.H. acknowledges her STFC studentship.
%%%%%%%%%%%%%%%%%%%%%%%%%%%%%%%%%%%%%%%%%%%%%%%%%%%%%%%%%%%%%%%%%%%%%%%%%%%%%%%%%%%%%%%
\bibliography{bibliography}
%\input{bibliography.bbl}
% Produces the bibliography via BibTeX.

\end{document}